# Probing Phonon Modes in Reconstructed twisted Homo and Hetero Bilayer System

Sushil Kumar Sahu[1†], Robin Bajaj[2†], Syed Ummair Ali[1†], Ajay Bhut[1], Roshan Jesus Mathew[1], Shinjan Mandal[2], Kenji Watanabe[3], Takashi Taniguchi[3], Manish Jain[2], and Chandan Kumar[1*]

[1]*Centre for Nano Science and Engineering, Indian Institute of Science, Bangalore 560012, India.*
[2]*Department of Physics, Indian Institute of Science, Bangalore 560012, India.*
[3]*National Institute for Materials Science, 1-1 Namiki, Tsukuba, 305-0044 Japan.*
† These authors contributed equally to the work
* Correspondence to: kchandan@iisc.ac.in

**Abstract**

**Twist-angle engineering in van der Waals homo and hetero-bilayers introduces profound modifications in their electronic, optical and mechanical properties due to lattice reconstruction. In these systems, the interlayer coupling and atomic rearrangement strongly depend on the twist angle, leading to the formation of periodic moiré superlattices. At small twist angles, significant lattice relaxation results in the emergence of domain structures separated by one-dimensional (1D) soliton networks, influencing electronic band structures and phonon modes. In this study, we systematically investigate the impact of lattice reconstruction on phonon renormalization in twisted bilayer graphene (TBLG) and graphene-hBN moiré superlattices, representing homo and hetero bilayer system, respectively. Using Raman spectroscopy, we identify distinct phonon behaviours across different twist angle regimes. In TBLG, we observe the evolution of the G peak, including broadening, splitting, and the emergence of additional peaks in the small angle range ($0.3°-1°$), attributed to moiré-modified phonon interactions. At large twist angles, the peaks gradually merge back into a single feature, reflecting the reduced impact of lattice reconstruction. Similarly, in hBN-graphene moiré superlattices, we detect moiré-induced Raman peaks above and below the G peak, while the central G peak remains largely invariant to twist angle variation. The theoretical calculations based on classical force-field uncover moiré phonon modes originating from different stacking regions, including AB (AB′), AA, and SP configurations, providing insights into phonon renormalization driven by lattice reconstruction. Our results establish a direct link between twist angle, lattice reconstruction, moiré phonons, and interlayer coupling, offering a fundamental framework for understanding phonon engineering in twisted bilayer systems. These findings pave the way for controlling phononic, optoelectronic and heat flow properties in next generation van der Waals heterostructures.**

**Introduction**

Stacking of atomic lattices with a relative twist angle leads to the formation of moiré pattern with twist angle dependent electronic[1–4], optical[5–7], and magnetic properties[8–10]. Twist angle, in addition to modifying the band structure also leads to substantial lattice reconstruction[11–13]. The lattice reconstruction in low angle hBN-graphene moiré superlattices, twisted bilayer graphene (TBLG) and twisted transitional metal dichalcogenide (tTMDC) has been shown experimentally using Transmission Electron Microscopy[11,12,14], scanning tunnelling microscopy[15], piezo response force microscopy[16], nano-infrared spectroscopy[17], Scanning Microwave Impedance Microscopy[18], nano-Raman[13] and atomic single electron transistor[19]. The lattice reconstruction in TBLG leads to a flat electronic band near magic angle[20–22], $\theta_c \sim 1.1°$. Moreover, in minimally TBLG ($\theta \ll \theta_c$) and tTMDC ($\theta < 2°$), a significant lattice reconstruction takes place leading to the formation of a narrow 1D solitons connecting energetically unfavourable AA stacking regions and separating favourable AB/BA Bernal stacking regions[5,23–28]. The reconstructed structure presents an opportunity for tuning the electronic and phonon dispersion[29–33].

Raman spectroscopy is a versatile tool to probe the phonon modes[34,35] that offers insights into the lattice dynamics, thereby providing an in-depth characterization of the homo[36–49] and hetero bilayer system[38,50–52]. In graphene (Gr)/hexagonal boron nitride (hBN) superlattice at non-zero twist angle additional peaks are observed above and below G peak which are typically associated with intralayer longitudinal optical (LO) and transverse optical (TO) modes, respectively which become Raman active by the moiré potential[38,51]. In addition, an increase in FWHM of the 2D peak is observed for $\theta \leq 2°$ which can be used to calculate the twist angle[51]. TBLG at higher angles ($\theta > 2°$) shows two peaks around the G peak near $1520\ cm^{-1}$ and $1625\ cm^{-1}$ which are associated to intralayer LO and interlayer TO modes, respectively[38]. Recently, the localization of phonon modes was shown in low angle ($\theta = 0.1°$) TBLG using nano-Raman technique[13]. However, a systematic study of moiré phonon with varying twist angle in reconstructed homo (Gr/Gr) and hetero (Gr/hBN) bilayer is still lacking.

In this study, we investigated how lattice reconstruction affects phonon renormalization in TBLG and Gr/hBN moiré superlattices, which represent homo- and hetero-bilayer systems, respectively. We find that the effect of lattice reconstruction on moiré phonon is quite distinct in homo and hetero bilayer system. For the homo-bilayer system (TBLG), we identified three twist angle regimes based on G peak Raman spectra: small ($\theta \leq 0.2°$), intermediate ($0.3° \leq \theta \leq 1°$), and large angles ($\theta \geq 2°$). In the small-angle regime, the G peak line shape of the Raman spectrum exhibits a single Lorentzian peak whose full width at half maximum (FWHM) increases with the twist angle. In the intermediate angle regime, we observe splitting of G peak ($G^+, G^-$) along with the appearance of two additional peaks - one above $G^+$ and one below $G^-$ peak. The splitting in G peak is observed till $\theta = 1°$. In the large angle regime, only one Lorentzian peak is observed which is independent of twist angle. Similarly in the case of hetero bilayer system we define four angle regimes – small ($\theta \leq 0.4°$), intermediate ($0.6° \leq \theta \leq 1.1°$), large ($1.4° \leq \theta \leq 1.9°$) and extra-large regime ($\theta \geq 2°$). The small and large angle regime displays a central G peak, and two additional peaks located above & below the central G peak, which we refer to as $M$ & $M'$. In the intermediate angle regime, $M$ & $M'$ peaks further split into two additional peaks. Thus, having in total five peaks in the intermediate angle regime. However, we do not find the splitting of central G peak with twist angle. In addition, in the extra-large angle regime, a single G peak is observed which is independent of twist angle. Our theoretical calculations, based on classical force fields, provide insight into the effect of the moiré superlattice on the splitting of the G mode. By examining various twist angles in minimally twisted Gr/Gr homo bilayers and Gr/hBN hetero bilayers, we attribute spectral peaks to distinct stacking regions, achieving results that align well with experimental observations. Our work provides a comprehensive study of moiré phonons and shows for the first time the intricate role of lattice reconstruction in the modulation of phonon dispersion at varying twist angles in homo and hetero bilayer graphene system.

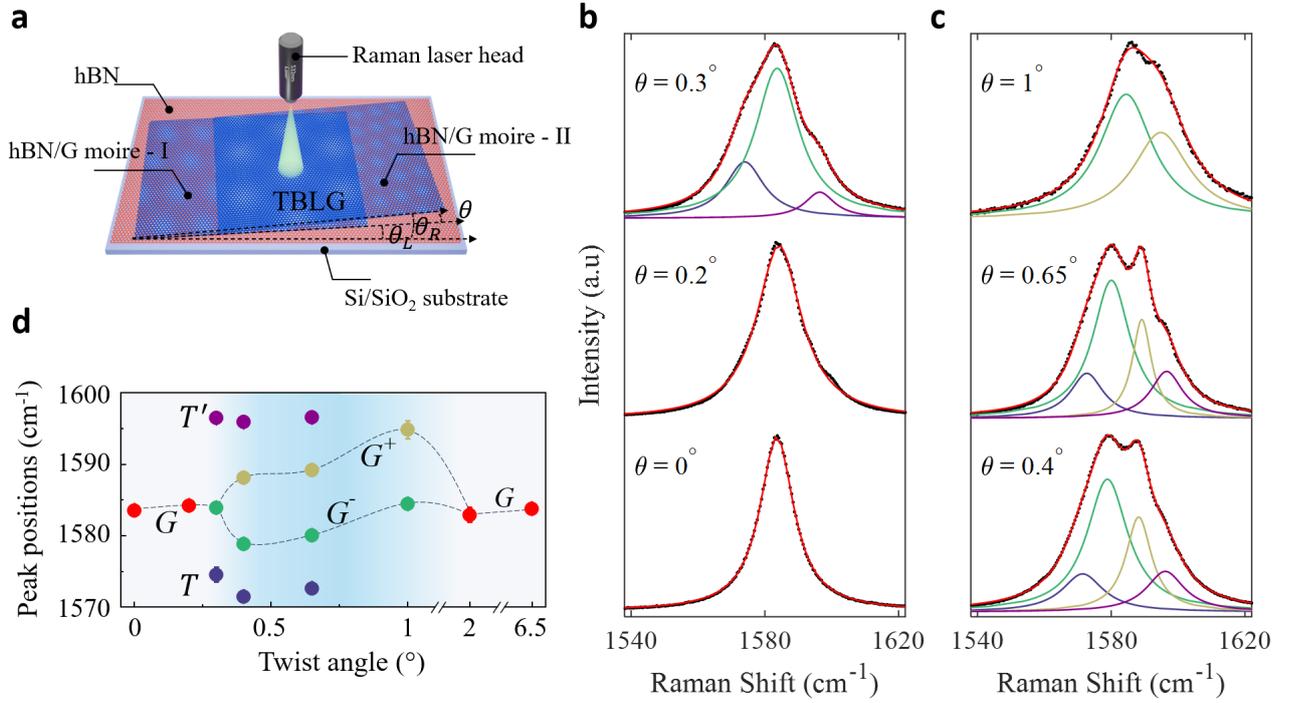

*Figure 1:* **Moiré Phonon in twisted bilayer graphene. a.** Schematic of a typical device. In our devices we align each graphene with the hBN to create hBN-graphene moiré superlattice. Moreover, we have also aligned two graphene with a small twist angle, forming TBLG. The twist angle of left ($\theta_L$) and right ($\theta_R$) hBN-graphene superlattice is determined by measuring the FWHM of the respective 2D peak. This device geometry allows us to measure the twist angle of TBLG which can be written as $\theta = |\theta_L - \theta_R|$. **b-c.** Raman spectra of twisted bilayer graphene (TBLG) at various twist angle. We find a clear evolution of G peak from a single Lorentzian peak at $\theta \sim 0°$ to four peak Lorentzian at intermediate angle and two peak Lorentzian at $\theta \sim 1°$. **d.** Peak positions as a function of twist angle. The peak positions are extracted from the fitted Lorentzian curves in Fig. 1b-c and the dashed lines are guide to the eye. A clear splitting of G peak is observed for $0.4° \leq \theta \leq 1°$.

## Raman Spectra of Twisted Bilayer Graphene

The schematic of the device is shown in Fig. 1a. Our devices consist of twisted bilayer graphene encapsulated between hexagonal boron nitride (hBN). The devices are fabricated using standard tear and stack method[1,53]. During the stacking process, we align both the graphene with the hBN. The Gr/hBN alignment leads to the formation of moiré superlattices. In addition, we also align two graphene edges with a small twist angle, forming TBLG (SI Section1). The angle of left ($\theta_L$) and right ($\theta_R$) Gr/hBN superlattices is determined by measuring the respective 2D peak FWHM. This device geometry allows us to measure the twist angle of TBLG which can be written as $\theta = |\theta_L - \theta_R|$. This unique device geometry provides a fast and easy way to determine the twist angle of TBLG at room temperature (SI section S3).

The Raman Spectra is obtained using a single excitation wavelength of 532 nm green laser with power less than 5 mW (SI section 2). Fig1 b-c shows the Raman spectra of TBLG. As expected at $\theta = 0°$, we see a usual G peak which can be fitted with a single Lorentzian peak. With increasing twist angle a gradual broadening of the peak is observed. At $\theta = 0.3°$, a further broadening in G peak can be observed. Moreover, at $\theta = 0.3°$ our data shows a clear hump and a shoulder above and below G peak, respectively. We find that three peak Lorentzian fitting provides a good fit to our full G peak spectrum. With the further increase in twist angle ($\theta = 0.4°$ and $0.65°$), along with the hump and a shoulder observed at $\theta = 0.3°$ we also find a clear splitting of G peak, which we label as $G^+$ and $G^-$. The full G peak spectrum can now be fitted with four peak Lorentzian. For $\theta = 1°$, we no longer observe the hump and shoulder around G peak. However, splitting of G peak can still be found and the spectrum can be fitted with two Lorentzian peaks. To further trace the peak evolution, we plot peak positions as a function of twist angle in Fig. 1d. The peak positions are extracted from the fitted Lorentzian curves in Fig. 1b-c and the dashed lines are guide to the eye. Based on the number of peaks in G peak full spectrum we can identify three angle regions – small angle ($\theta \leq 0.2°$), intermediate angle ($0.2° < \theta < 1°$), and large angle ($\theta \geq 2°$). As can be seen from the figure, a single G peak in small angle regime splits into $G^+$ and $G^-$ along with the additional new peaks above $G^+$ and below $G^-$ in the intermediate angle regime. Finally in the large angle regime $G^+$ and $G^-$ seems to merge again and a single G peak is obtained.

To understand the splitting of the G peak in twisted bilayer graphene at different twist angles, we theoretically investigate the moiré phonon modes at the Γ point in the moiré Brillouin zone using the PARPHOM[54] package. The moiré phonon eigenvalues and eigenvectors are computed for a relaxed moiré structure by diagonalizing the dynamical matrix constructed from force constants. The relaxation of the structure and generation of force constants is done using classical force fields within the LAMMPS[55] package (See SI Section S6 for details of the calculation).

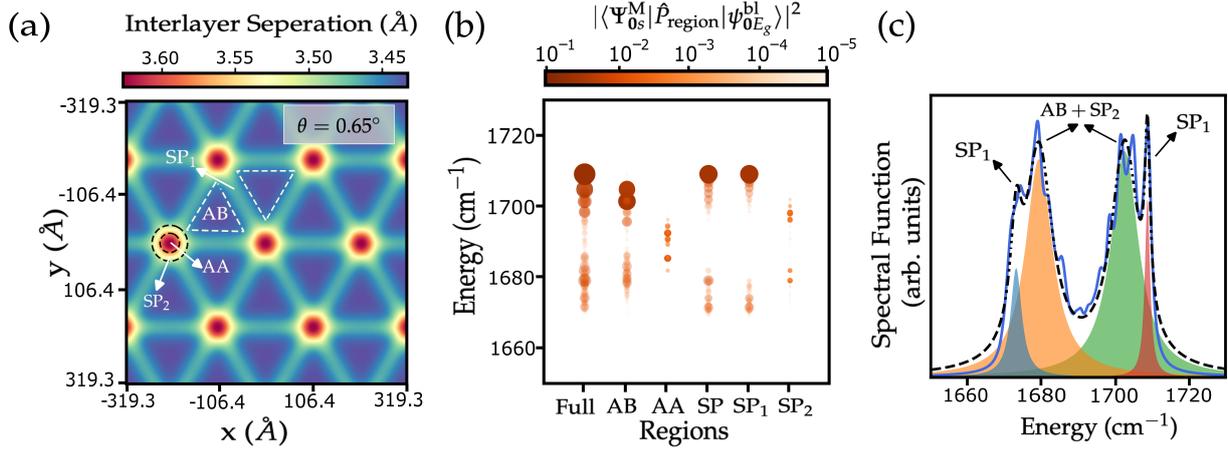

*Figure 2:* **Theoretical calculations of Moiré Phonons in twisted bilayer graphene ($\theta \sim 0.65°$). a.** Interlayer separation landscape (ILS) of $\theta \sim 0.65°$ TBLG including lattice relaxation effects with color bar showing the range of interlayer separation distances in Angstroms. The stacking regions AB, AA, $SP_1$ and $SP_2$ have been marked. **b.** Total and region-wise projection weights of the moiré modes onto the Gr/Gr bilayer-like $E_g$ mode as a function of energy of the moiré phonon modes for the regions AB, AA, SP, $SP_1$ and $SP_2$ (shown as insets) classified based on ILS in 2a. The colorbar shows the squared modulus of the projection weights. **c.** Spectral function as a function of energy with the theoretically calculated total spectra (shown with blue curve) and the best fit to the spectra using the four Lorentzians with the peak positions at 1673, 1679, 1702, and 1709 cm$^{-1}$ (shown with black dotted curve). The peaks have been attributed to the corresponding stacking regions and marked with arrows.

Utilizing the theoretical framework described above, the dependence of projection weights on θ and the contribution from various stacking regions at higher angles (1° − 2°) have already been studied in Ref[56]. In this regime, $E_g$ splits into two peaks. The high-frequency and lower-frequency peaks are attributed to the combination of AB, $SP_1$, and $SP_2$ regions, with AB and $SP_1$ significantly contributing to the high-frequency peaks. Different regions are shown in Fig. 2a and explained in the following sections. For $\theta > 2°$, as the moiré potential weakens and the two layers become uncorrelated, a single peak emerges.

To elucidate the origin of this quadruplet splitting observed within the range $0.4° < \theta < 0.65°$, we employ the same theoretical framework. For this investigation, we specifically consider Gr/Gr twisted at 0.65° as the system under study (See SI Section 6 for calculation details). We first calculate the interlayer spacing landscape for a Gr/Gr homo bilayer at $0.65°$ and classify the stacking regions into four types: AB, AA, $SP_1$, and $SP_2$, as shown in Fig. 2a. In the AB region, carbon atoms from one sublattice align directly with carbon atoms of the opposite layer, while the carbon atoms of the other sublattice align with the center of the hexagon formed by carbon atoms in the second layer. In the AA region, both carbon atoms of one layer sit directly above the corresponding carbon atoms of the second layer. The SP region, which lies between AA and AB, is further divided into two subregions: $SP_1$ and $SP_2$. These two subregions differ in interlayer spacing and are highlighted with different colors in Fig. 2a.

We then calculated the projection weights of the moiré phonon modes onto the bilayer-like G phonon mode, defined analogously to the G mode defined in Ref[56], for all atoms in the moiré unit cell. The region-specific projection weights were determined for AB, AA, SP ($SP_1 + SP_2$), $SP_1$, and $SP_2$ regions and are shown in Fig. 2b. Our results show that the prominent contributions to the projection weights corresponding to frequencies around 1673, 1679, 1702, and 1709 cm$^{-1}$ are primarily from $SP_1$, AB + $SP_2$, AB + $SP_2$, and $SP_1$ regions, respectively. The projection weights for the AA region are negligibly small due to its minimal size at small twist angles, contributing insignificantly to the overall projection weights. This clearly explains the observed splitting of the G peak into four distinct peaks at a twist angle of 0.65°, while highlighting the contributions of each stacking region.

To directly compare the calculated moiré phonon eigenvalues with the peak positions of the experimental Raman spectra, we define a spectral function (including a phenomenological linewidths) as follows:

$$A_{0E_g}(\omega) = \frac{1}{\pi} \sum_\nu \frac{\left|\left\langle \Psi^M_{0\nu} \middle| \Psi^M_{0E_g} \right\rangle\right|^2 \Gamma^{0\nu}}{(\omega - \omega_{0\nu})^2 + (\Gamma^{0\nu})^2}$$

where $\Psi^M_{0\nu}$ is the eigenvector of the $\nu^{th}$ moiré phonon mode at the Γ point of moiré unit cell, $\Psi^{bl}_{0E_g}$ is the eigenvector of bilayer-like G mode in the moiré unit cell, $\omega_{0\nu}$ are the moiré phonon mode frequencies, and $\Gamma^{0\nu}$ are the electron-phonon linewidths which we take phenomenologically as 1 cm$^{-1}$. The calculated spectral function is shown in Fig. 2c. Fitting four Lorentzians to this spectral function yields peak positions at 1673, 1679, 1702, and 1709 cm$^{-1}$. While these positions are slightly overestimated due to limitations in the classical force field for describing the optical modes, the model accurately captures the qualitative splitting of the G peak in intermediate angle regime.

To explain the evolution from four peaks at $0.65°$ to three peaks at $\theta = 0.3°$, it is necessary to perform similar calculations for moiré phonon modes at these smaller angles. However, this approach is computationally expensive due to the large number of atoms involved. Instead, we analyze the evolution of the interlayer spacing landscapes as a function of twist angle (See SI section 7 for more information). As the angle decreases from $0.65°$ to $0.2°$, the fraction of atoms in $SP_2$ and AA regions compared to fraction of atoms in AB region in the ILS landscapes decrease significantly, while the fraction of atoms in $SP_1$ region also decreases but remains notable compared to $SP_2$. This transformation explains the emergence of a three-peak structure at $0.3°$, where the second peak is predominantly from the AB region due to the disappearance of the $SP_2$ region, which previously caused peak splitting at higher angles. The first and third peaks remain associated with the $SP_1$ region. As the angle further decreases, the $SP_1$ region vanishes entirely, leaving the homo bilayer with a purely AB stacking configuration at $\theta = 0°$. This results in a single peak originating solely from the AB region. These results presents a coherent explanation for the experimental data illustrated in Fig. 1.

**Raman Spectra of Graphene/hBN moiré Superlattices**

Now we turn to Gr/hBN moiré superlattice, a hetero bilayer system. Fig. 3a-c plots the Raman spectra at various Gr/hBN twist angles ranging from $0°$ to $2°$. Similar to TBLG discussed above, based on the number of peaks in full G peak spectra we define three angle regime- small, intermediate & large. In the small angle regime, Fig. 3a, ($0° \leq \theta \leq 0.4°$), we find two additional peaks along with the central G peak. The new peaks are located above and below the G peak. In the intermediate angle regime ($0.6° \leq \theta \leq 1.1°$), we find emergence of four new peaks along with the central G peak. The two peaks are positioned above G peak while another two peaks lie below G peak. With the additional increase in angle ($1.4° \leq \theta \leq 2°$), we again find two peaks along with central G peak, similar to small angle regime. For devices with very large angle ($\theta > 2°$) we only find a single G peak.

To further understand the evolution of peaks as a function of Gr/hBN twist angle we plot in Fig. 3d peak position as a function of Gr/hBN twist angle. The peak position is extracted by fitting G peak Raman spectra. We find that G peak position does not seem to change with twist angle, this contrasts with TBLG where we found clear splitting of G peak with twist angle. Moreover, the new peaks ($M$ & $M'$) in small angle regime seems to split with increasing twist angle and recombines in large angle regime.

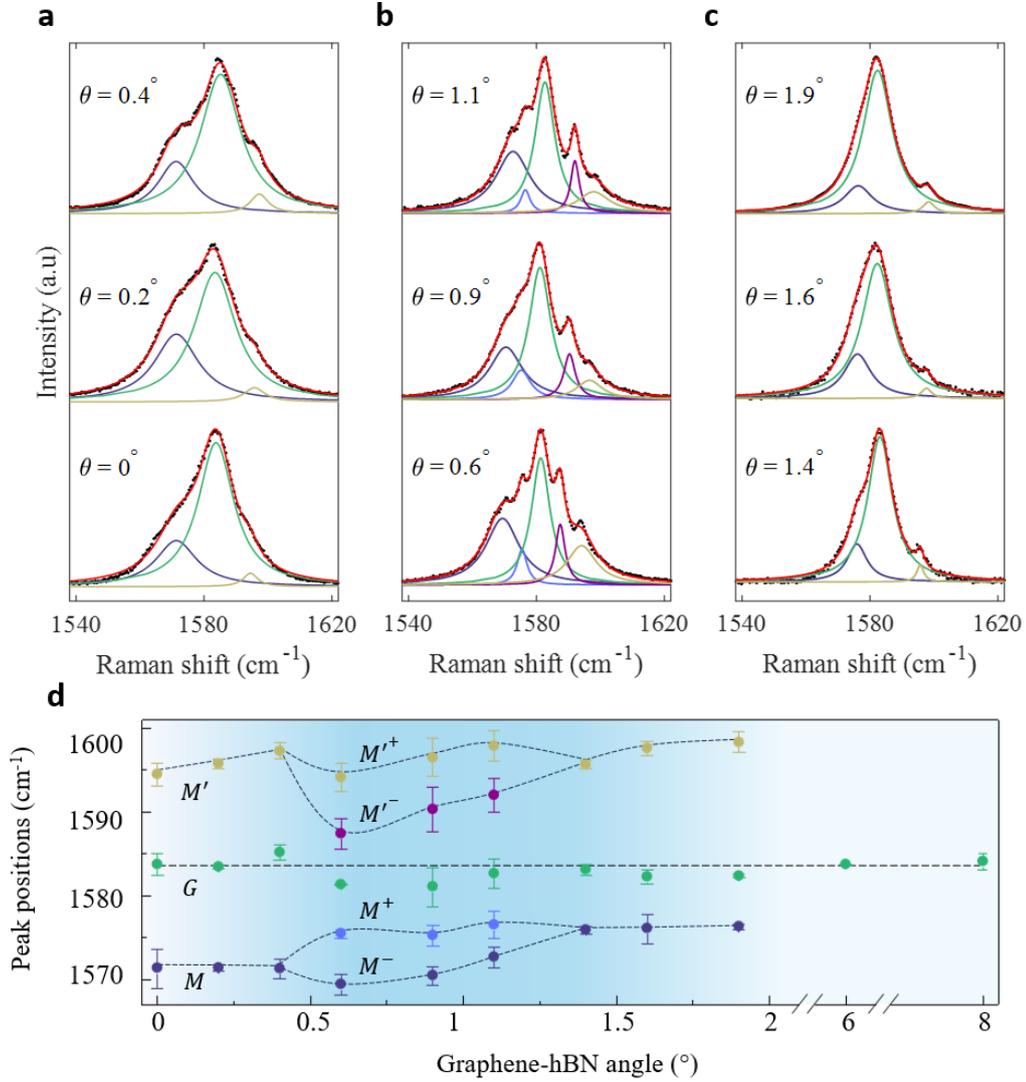

*Figure 3:* **Moiré Phonon in hetero bilayer**. **a-c.** Raman spectra of hBN-graphene moiré superlattice at various twist angle. We find splitting of G peak into three peaks at small angle ($0° \leq \theta \leq 0.4°$). At an intermediate angle ($0.6° \leq \theta \leq 1.1°$), we find five peaks within the G peak, two above the central G peak and two below it. Further increase in angle, ($1.4° \leq \theta \leq 1.9°$), we recover three peak splitting scenarios observed at low angle region. **d.** Evolution of peaks with hBN-graphene twist angle. We find that G peak position remains immune to twist angle variation. However, the $M$ and $M'$ peaks split into two at intermediate angle. At very large angle($\theta \geq 2°$) only G peak is observed.

To investigate the moiré phonons in Gr/hBN hetero bilayers, we applied the same theoretical framework used for Gr/Gr homo bilayers, calculating the moiré phonon modes for twist angles ranging from $0°$ to $12°$. A notable distinction lies in defining the bilayer-like $E_g$ mode for the Gr/hBN unit cell, which involves matching the lattice constants of graphene and hBN (refer to the SI Section 6 for further details).

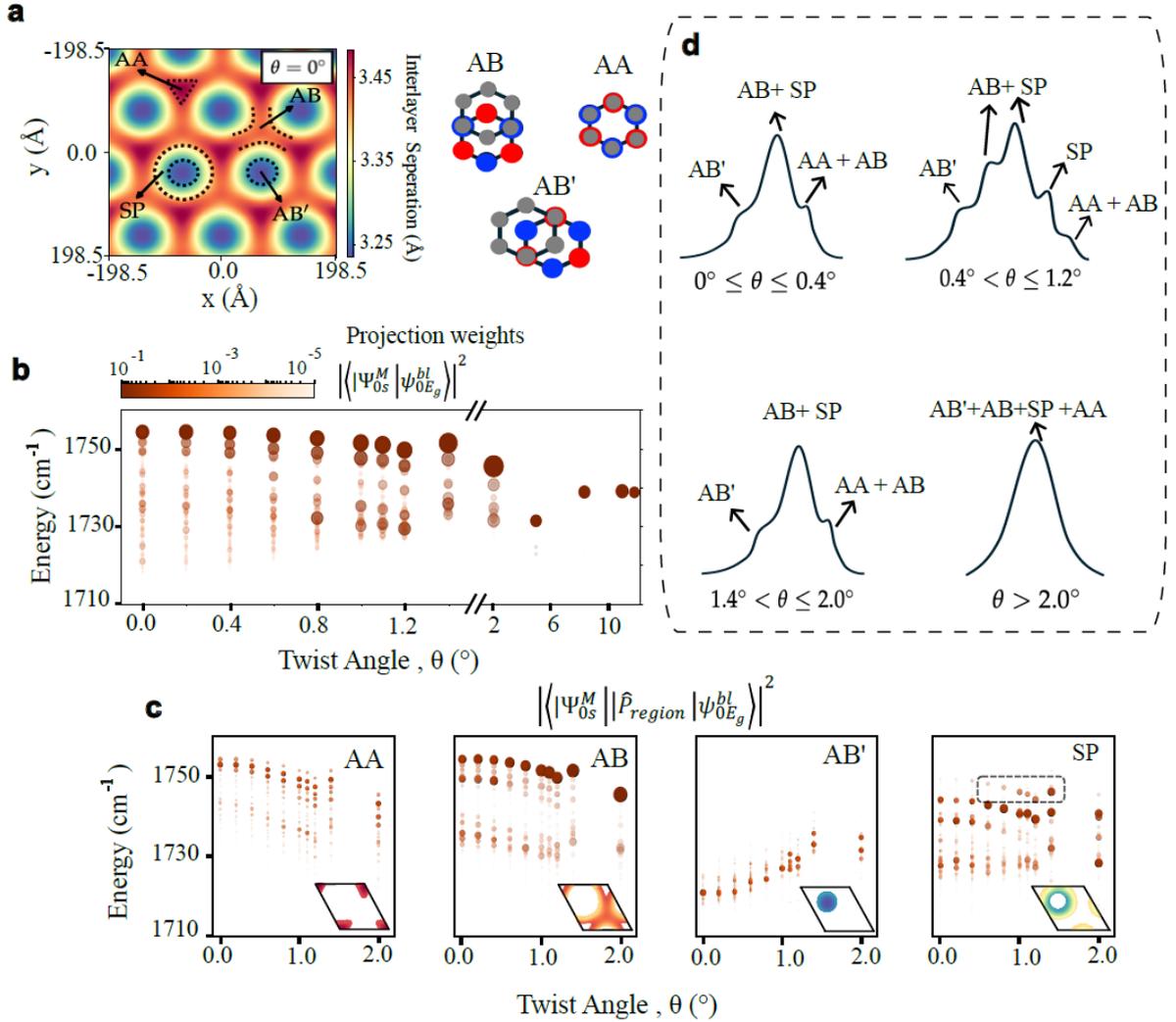

*Figure 4:* **Theoretical calculations of Moiré Phonons in Gr/hBN hetero bilayer. a.** Interlayer separation landscape (ILS) of $\theta = 0°$ Gr/hBN including lattice relaxation effects. The colour bar shows the range of interlayer separation distances in Angstroms. The stacking regions AB, AB', AA and SP have been marked in the ILS. The actual stacking of C, B and N atoms for the regions AB, AB' and SP has been shown. **b.** Total projection weights of the moiré modes onto the Gr/hBN bilayer-like $E_g$ mode for twisted Gr/hBN at the angles ranging from 0° to 12°. **c.** Region-wise projection weights as a function of twist 0° to 2° for the regions AA, AB, AB' and SP, respectively (shown as insets for $\theta = 0^0$) classified based on ILS in 4a. The colour bar shows the squared modulus of the projection weights in (4b, 4c). **d.** Schematic of contributions of different stacking regions to the peaks observed in different angle regimes.

Unlike Gr/Gr, where there are three main stacking regions (AA, AB, and SP), the presence of boron and nitrogen atoms in the hBN sublattices allows for further classification of the AB regions into AB and AB′. In AB stacking, the hBN layer is shifted such that the carbon atoms in one graphene sublattice align with boron atoms, while nitrogen atoms align with the centres of carbon hexagons. In AB′ stacking, the nitrogen atoms sit on top of carbon and boron atoms are at the centres of carbon hexagons (Fig 4a right panel). In the AA stacking, each graphene atom aligns directly with a corresponding boron or nitrogen atom in the BN layer. The SP region represents a distinct stacking configuration, separated by domain boundaries from the AA, AB, and AB′ regions. This spatial stacking pattern (SP) typically induces variations in electronic and vibrational properties. The stacking regions are defined based on the interlayer spacing landscape between graphene and the BN layer (Fig. 4a).

We start by calculating the full projection weights (Fig. 4b) by projecting the moiré phonon modes onto the bilayer-like $E_g$ mode of the Gr/hBN moiré unit cell for twist angles between $0°$ and $12°$ (refer to the Supplementary Information for further details). This analysis reveals substantial splitting of the $E_g$ mode for minimal twist angles due to the strong moiré potential. For twist angles in the range of $0.8°$ to $1.2°$, significant projection weights occur at different frequencies, potentially contributing to distinct spectral peaks as observed in experiments.

To gain a more comprehensive understanding, we computed region-wise projection weights for the four regions: AA, AB, AB′, and SP, across the entire range of twist angles. For the small angle ($0° \leq \theta \leq 0.4°$) region, we find that the high-frequency peak observed in experimental spectra consistently originates partially from both AA and AB regions for all twist angles, appearing near 1750 cm$^{-1}$. This systematic trend closely aligns with experimental observations, allowing us to associate the high-frequency peak with the AA+AB regions. The low-frequency peak in the spectra is clearly attributable to the AB′ region, appearing around 1720 cm$^{-1}$ for $\theta = 0°$. The position of this peak increases monotonically with twist angle as observed in experiment (Fig. 3d) and it gradually converges toward the G peak position, indicating its emergence from the AB′ region across all twist angles.

For higher twist angles ($\theta \geq 4°$), a middle G peak appears near 1735 cm$^{-1}$, with significant projection weights from both the AB and SP regions for twist angles within $\theta \leq 0.4°$ and $1.4° \leq \theta \leq 2°$. Thus, this middle G peak can be attributed to contributions from the AB+SP regions. The three-peaked spectra observed in the range of $\theta \leq 0.4°$ and $1.4° \leq \theta \leq 2°$ can be explained by this attribution. At twist angles $\theta \geq 2°$, a single peak emerges, resulting from contributions from all stacking regions (AA+AB+ AB′+SP) due to the weakening of the moiré potential and reduced layer correlation.

In the intermediate twist range ($0.4° \leq \theta \leq 1.2°$), additional spectral peaks appear. Region-wise projections for the SP region (See dotted box marked in the SP region of Fig. 4c) indicate emerging weights around 1745 cm$^{-1}$ exclusively within this angle range, which disappear at other twist angles. Therefore, the fourth peak can be completely attributed to the SP region. Furthermore, weights from the AB and SP regions are distributed over multiple frequencies within the 1725-1735 cm$^{-1}$ range, allowing the second peak to be associated with a combination of AB and SP regions. The five-peaked spectra observed in the range of $0.4° \leq \theta \leq 1.2°$ can be explained by this attribution. To visualize the region-wise attribution of various peaks across the entire range of twist angles, a schematic is presented in Fig. 4d. This illustration clearly demonstrates the alignment between experimental observations and the theoretical calculations of moiré phonons in heterostructures based on force field models.

**Summary**

Our work investigates the effect of lattice reconstruction on phonon renormalization in twisted homo and hetero bilayer systems, focusing on twisted bilayer graphene (TBLG) and hexagonal boron nitride (hBN)-graphene moiré superlattices. In TBLG, distinct behaviours in the Raman G peak were observed across different twist angle regimes, including the splitting of the G peak and appearance of new peaks at intermediate twist angles. For the graphene/hBN hetero-bilayer, four twist angle regimes were identified, reflecting the influence of stacking configurations such as AB, AB', and SP on the phonon spectrum. The results highlight how lattice reconstruction significantly influences the phonon modes and shows that phonon properties in these systems are highly sensitive to the twist angle, offering potential for tuning electronic and vibrational properties. This work contributes to the understanding of phonon dynamics in moiré superlattices and sets the stage for further research in phonon engineering.

**Acknowledgements:** This research was funded by an IISc start-up grant, Lakshmi Narayanan Young Investigator Grant, SERB core research grant (CRG/2023/006376). We thank the CeNSE facilities for their support, which is funded by the Ministry of Human Resource Development (MHRD), the Ministry of Electronics and Information Technology (MeitY) and the Department of Science and Technology (DST). M.J. acknowledges the Nano Mission of the Department of Science and Technology for financial support under Grant No. DST/NM/TUE/QM-10/2019. R.B. acknowledge the funding from the Prime Minister's research fellowship (PMRF), MHRD. We thank S. Ilani and V. Meunier for useful discussion.

**Contributions:** C.K conceived the project and designed the experiments. S.U.A, R.J.M., S.K.S performed the experiments. S.U.A, S.K.S, A.B fabricated the devices. S.U.A, S.K.S and C.K analysed the data. R.B, S.M and M.J wrote the theoretical model. K.W and T.T supplied the hBN crystals. S.U.A, S.K.S, R.B, M.J, and C.K wrote the manuscript with input from other authors.

**Data availability:** The data that support the plots and other analysis in this work are available from the corresponding author upon request.